\documentclass[twocolumn]{aastex631}
\usepackage{CJK}

\usepackage[T1]{fontenc}

\usepackage{newtxtext,newtxmath}
\usepackage[figure,figure*]{hypcap}
\usepackage{booktabs}
\usepackage{cleveref}
\usepackage{paralist}
\graphicspath{{fig/}}
\usepackage{ulem}
\usepackage{natbib}
\usepackage[shortlabels]{enumitem}

\usepackage{xcolor,colortbl}  

\usepackage[colorlinks=true]{hyperref}

\begin{document}
\begin{CJK*}{UTF8}{gbsn}

\title{On the structure of open clusters: geometric vs geomantic}


\email{lilu@shao.ac.cn, zyshao@shao.ac.cn}

\author[0000-0002-0880-3380]{Lu Li (李璐)}
\affil{Shanghai Astronomical Observatory, Chinese Academy of Sciences, 80 Nandan Road, Shanghai 200030, China.}

\author[0000-0001-8611-2465]{Zhengyi Shao（邵正义）}
\affil{Shanghai Astronomical Observatory, Chinese Academy of Sciences, 80 Nandan Road, Shanghai 200030, China.}
\affil{Key Lab for Astrophysics, Shanghai 200234, China}

\begin{abstract}

Understanding our place in the universe is an eternal quest. Through the analysis of the 3D structures of 66 nearby open clusters using Gaia DR3 data, we discovered an intriguing pattern: most clusters show their elongation directions pointing at the Sun, suggesting that the Solar System might just be the universe's favorite spot, a cosmic feng shui hotspot! This surprising result hints at a subtle blend of geometry and geomancy. 

\end{abstract}
 

\section{Introduction} \label{sec:intro}

Since ancient times, humans have sought to understand their place in the universe. Ancient civilizations across the world, from the Greeks to the Chinese, believed that the Earth was undoubtedly the cosmic center, with everything in the sky harmoniously revolving around us. This geocentric idea wasn't just science. It was a worldview rooted in the idea of cosmic harmony and even a touch of geomancy, or feng shui in Chinese culture, the art of finding the best spot in space. 

Then came Copernicus, moving the center from Earth to the Sun, and later, modern cosmology declared that the universe has no unique center. The cosmological principle tells us that the universe is homogeneous and isotropic on large scales, no special place, no favored position.

But what if the Solar System, despite all modern ideas, actually holds a special place in the cosmic map? Could there be a subtle hint of cosmic feng shui where the universe itself shows preference for the Sun's position? Perhaps the universe itself has a preference for where our star happens to be an optimal cosmic location where everything just lines up nicely. 

To celebrate April Fool's Day, we decided to put this bold idea to the test by examining the three-dimensional structure of open clusters (OCs) in the Solar neighborhood. Using data from Gaia DR3, we analyzed the elongation and alignment of 66 OCs within 500 pc to investigate whether the Solar System holds a privileged spatial position in the local universe.

\begin{figure*}[!htbp]\label{fig:oc_struct}
  \centering
  \includegraphics[width=1\textwidth]{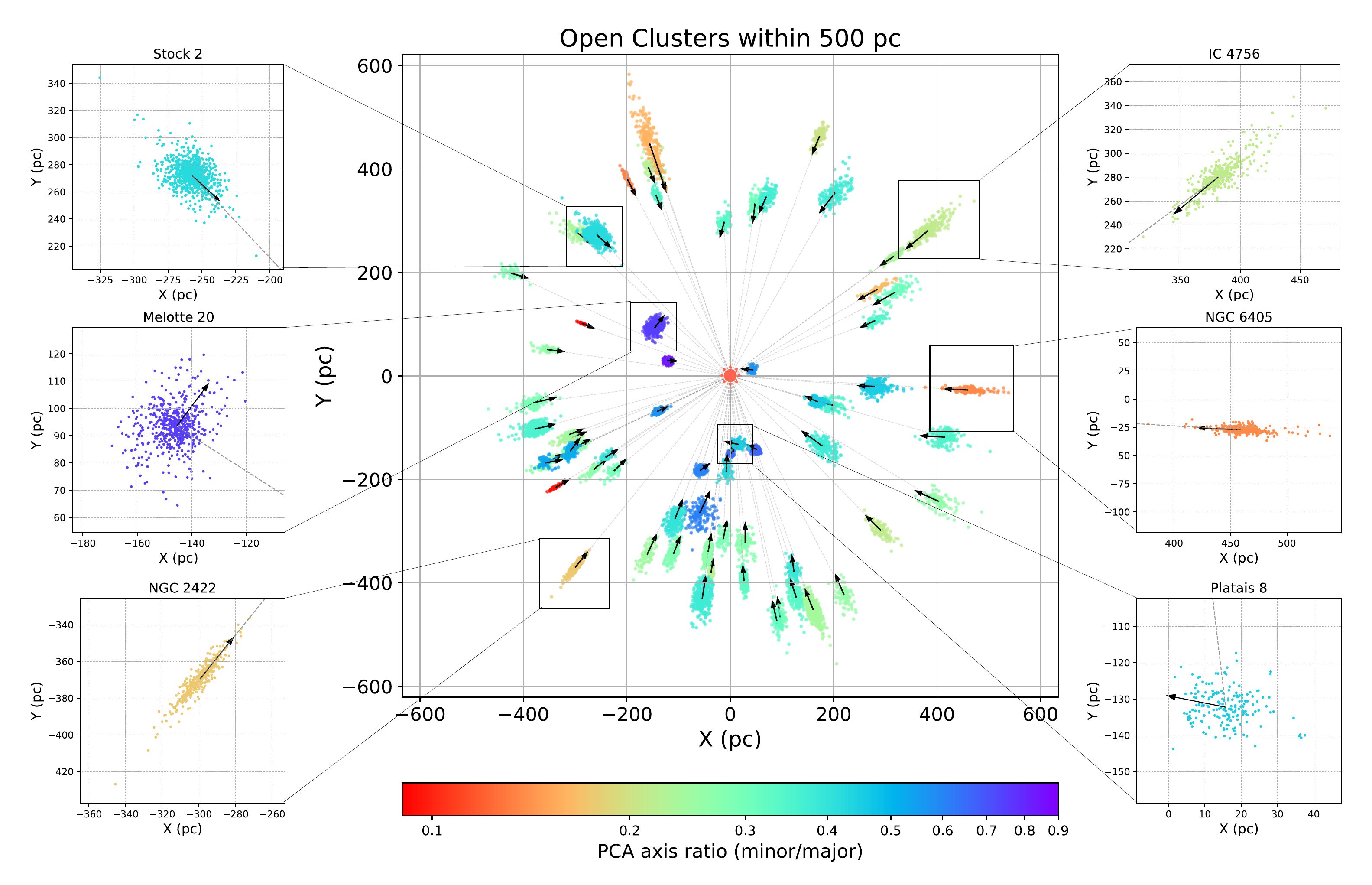}
  \caption{The distribution of the 66 selected open clusters in the Galactic Cartesian coordinate. The main panel shows the spatial distribution of all clusters, with colors representing their axis ratio. The black arrows are the direction of the major axis as determined by PCA, and the dotted gray line denotes the direction toward the Solar System. The zoomed-in panels highlight representative clusters chosen to cover a range of axis ratios and alignment angles, demonstrating different levels of elongation and spatial orientation.}
\end{figure*}

\section{Data}

To investigate the three-dimensional structure and spatial orientation of nearby OCs, we selected clusters from the \citet{2020A&A...640A...1C} catalog, which provides reliable cluster membership data derived from Gaia astrometric data. The selection criteria were as follows:

\begin{enumerate}
    \item \textbf{Distance Criterion:} We selected clusters within a 500 pc radius from the Solar System to ensure high astrometric precision and minimize parallax errors.
    \item \textbf{Number of Members Criterion:} Only clusters with more than 100 member stars were chosen to guarantee reliable structure determination.
\end{enumerate}

After applying these selection criteria, a total of 66 OCs were identified and retained for further analysis. We utilized the astrometric data (Right Ascension, Declination, and parallax) of member stars from Gaia DR3 \citep{2023A&A...674A...1G} and transformed these coordinates into a Galactic Cartesian coordinate system centered at the Solar System. 

\section{Method}

To analyze the three-dimensional structure of each OC, we employed Principal Component Analysis (PCA), a statistical technique that identifies the principal axes of variance within a dataset. PCA allows us to quantify the spatial distribution of cluster members and determine the primary directions of elongation. By using the three-dimensional Cartesian coordinates \((X, Y, Z)\) of member stars obtained from Gaia DR3, we decomposed the variance into three orthogonal components corresponding to the principal axes.

The principal axis associated with the \textbf{largest variance} represents the \textbf{major axis} of the cluster, while the axis associated with the \textbf{smallest variance} corresponds to the \textbf{minor axis}. The ratio of the standard deviations along the minor and major axes quantifies the degree of structural elongation. An axis ratio significantly less than 1 indicates a highly elongated structure, while a ratio closer to 1 suggests a more spherical configuration.

We also investigate the spatial alignment of the cluster structures relative to the Solar System, we computed the alignment angle between the major axis of the cluster and the line connecting the cluster centroid to the Solar System. This angle quantifies whether the elongation direction points towards the Sun. Smaller angles indicate stronger alignment between the cluster's elongation direction and the Cluster-Sun line.

\section{Results}

Our analysis of the 66 selected OCs reveals a wide range in their structural elongations. The axis ratios span from 0.05 for highly elongated clusters to around 0.87 for nearly spherical. Figure \ref{fig:oc_struct} illustrates the X -- Y distribution for all the selected OCs. 

We can see that the PCA decomposition confirms that each cluster possesses a well-defined major axis along which the stellar distribution is maximized. The direction of this major axis is shown by black arrows in Figure \ref{fig:oc_struct}. The dotted grey line represents the direction toward the Solar System. A visual inspection reveals that a considerable number of clusters have their major axes oriented toward or near the Solar direction, hinting at a possible preferential alignment.

The zoomed-in panels in Figure \ref{fig:oc_struct} highlight representative clusters chosen to cover a range of axis ratios and alignment angles, showcasing different levels of elongation and spatial orientation. While some clusters exhibit a strong alignment with the Solar direction, others appear randomly oriented. Interestingly, these randomly oriented clusters tend to have a more diffuse spatial distribution, suggesting that they may be undergoing dynamical processes such as tidal disruption.

\section{Conclusion}
We investigated the three-dimensional structure and spatial orientation of 66 OCs within 500 pc using Gaia DR3 data. PCA revealed a broad range of structural elongations, with most of the OCs appearing preferentially aligned with the Solar direction.  

Our analysis has revealed a shocking truth: the Solar System really is a special place! Most of the OCs we studied have their elongation directions pointing straight at us, as if the universe itself has decided that the Sun is the ultimate cosmic feng shui hot-spot. This work echoes the modern cosmological principle: where the observer stands, there is the center of the universe.

\section{Disclaimer}
It is important to clarify that the apparent elongation of open clusters observed in this study does not necessarily correspond to true physical elongation. Instead, this effect arises primarily from the inherent uncertainties in parallax measurements, which are generally larger than those in right ascension and declination. As a result, the calculated elongation may be exaggerated along the line of sight, giving the illusion of an elongated structure. We strongly suggest that researchers claiming to discover elongation effects along the line of sight in three-dimensional structures of open clusters carefully consider the impact of measurement uncertainties.

\section*{acknowledgments}

LL thanks Dr. Zhaozhou Li for the insightful discussion and Prof. Shiyin Shen for the peer pressure that motivated us to complete this paper.

\vspace{5mm}

\software{astropy \citep{2013A&A...558A..33A}}

\bibliography{main}
\bibliographystyle{aasjournal}

\end{CJK*}
\end{document}